\documentclass[aps,twocolumn,amsmath,amssymb,floatfix,showpacs]{revtex4}

\usepackage{graphicx}
\usepackage{dcolumn} 
\usepackage{bm}      
\usepackage{color}

\begin{document}

\title{Experimental evidence of conformal invariance in soap film turbulent flows}
\author{S.~Thalabard$^1$, M.I.~Auliel$^2$, G.~Artana$^2$, P.D.~Mininni$^{3,1}$, and 
        A.~Pouquet$^1$}
\affiliation{$^1$Computational and Information Systems Laboratory, NCAR, 
         P.O. Box 3000, Boulder, Colorado 80307-3000, USA. \\
             $^2$Laboratory of Fluid Dynamics, Facultad de Ingenier\'{\i}a, Universidad 
         de Buenos Aires and CONICET, Paseo Col\'on 850, C1063ACV Buenos Aires, 
         Argentina. \\
             $^3$Departamento de F\'\i sica, Facultad de Ciencias Exactas y
         Naturales, Universidad de Buenos Aires and IFIBA, CONICET, Ciudad 
         Universitaria, 1428 Buenos Aires, Argentina.}
\date{\today}

\begin{abstract}
We present experimental evidence of statistical conformal invariance in isocontours of 
fluid thickness in experiments of two-dimensional turbulence using soap films. A Schlieren 
technique is used to visualize regions of the flow with constant film thickness, and 
association of isocontours with Schramm-L\"owner evolution (SLE) is used to identify 
conformal invariance. In experiments where an inverse energy cascade develops, statistical 
evidence is consistent with such an association. The diffusivity of the associated 
one-dimensional Brownian process is close to $8/3$, a value previously identified 
in isocontours of vorticity in high-resolution numerical simulations of two-dimensional 
turbulence (D.~Bernard et al., Nature Phys. 2, 124, 2006). In experiments where the inverse 
energy cascade is not sufficiently developed, no statistical evidence of conformal 
invariance is found.
\end{abstract}
\maketitle

Two-dimensional (2D) turbulence displays important differences with the three-dimensional 
(3D) case \cite{Kraichnan80}. While turbulent flows in 3D tend to create smaller 
scales and disorder, flows in 2D tend to self-organize and create long-living coherent 
structures \cite{Kraichnan67}. These differences are not just of academic interest, as 
many flows in nature are quasi-2D. This is often the result of volume forces that impose a 
preferred direction, as e.g., rotation and/or stratification in the atmosphere, and magnetic 
fields in the interplanetary medium. In particular, the atmosphere can be considered to a 
good degree of approximation as a shallow layer of fluid, as most of the weather takes 
place in a thin layer (the troposphere) of approximately 10 km depth, while the larger 
horizontal scales are of the order of thousand kilometers \cite{Pedlosky}.

2D turbulence is far from being completely understood. If injected at intermediate scales, 
enstrophy (the squared vorticity) is transferred towards smaller scales, in a so-called 
{\it direct cascade}, while energy is transferred towards larger scales in an 
{\it inverse cascade} \cite{Kraichnan67,Kraichnan80}. Besides the direction of the cascade, 
and unlike its 3D counterpart, the energy cascade is scale invariant: while in 3D turbulence 
strong intermittent events develop that make the flow multi-fractal, 2D flows are absent of 
such intermittency at large scale \cite{Siggia81,Paret98}.

An even stronger form of symmetry was identified in recent years: that of conformal 
invariance \cite{Bernard06}. While scale-invariance is a global property, often quantified 
using two-point correlation functions, conformal invariance is a local property involving 
invariance under conformal (angle-preserving) transformations. Conformal invariance 
is used in quantum field theories, including applications in condensed matter 
\cite{Belavin84}. In fluid dynamics, it is often used to solve Laplace equation for ideal 
incompressible and irrotational flows in 2D domains. That 2D turbulence (a much harder 
problem where non-linear effects are dominant) may be conformal invariant was originally 
proposed in \cite{Polyakov93}. Its confirmation requires measurement of multi-point 
correlation functions. Instead, in a recent work \cite{Bernard06}, evidence of conformal 
invariance in numerical simulations of 2D turbulence was shown using Schramm-L\"owner 
evolution (SLE) \cite{Lowner23}. SLE applies to a special set of random curves without 
self-intersections in a 2D space, and it was shown that if the conformal transformation 
that maps curves in the 2D space into the real axis gives a one-dimensional (1D) Brownian 
process (BP), then the original curves are statistically conformal invariant. This result was 
used to show that isocontours of zero vorticity in simulations of 2D turbulence are 
conformal invariant (and moreover, they are generated by an SLE process) \cite{Bernard06}. 
The analysis also allowed identification of the class of universality to which these curves 
belonged, identified by the diffusivity $\kappa$ of the BP, and noted $SLE_{\kappa}$. These 
classes share common statistical properties, such as critical exponents. More recently, the 
result was used to show that isocontours of temperature in a quasi-geostrophic model also 
were SLE, with a different $\kappa$ \cite{Bernard07}.

It is unclear however how these results can be extended to more realistic configurations. 
Under favorable conditions, flows in nature can be considered as quasi-2D at best. They 
are not exactly 2D as they may have small (but finite) thickness, or translation symmetry 
in the out-of-plane direction. As a consequence, motions outside the plane of interest can 
develop. Are conformal invariant properties still relevant in such cases? Moreover, the 
method used to detect conformal invariance in 2D turbulence relies on the vorticity, a 
quantity which is hard to measure with small uncertainties in experiments or observations. 
In many cases, vorticity is obtained after applying a curl to particle image velocimetry 
(PIV) measurements. This results in a loss of spatial resolution (PIV gives, as a rule, 
the velocity field every six or more image pixels, per virtue of the interrogation window 
used for the correlation), and in an increase of error per virtue of the numerical 
derivatives required to compute the curl. In some cases, spatial resolution of the 
correlation technique can be improved by using flow optic approaches with PIV images. 
However, an artificial smoothing of the results may still take place as a consequence of 
the regularization employed.

\begin{figure}
\includegraphics[width=4.05cm]{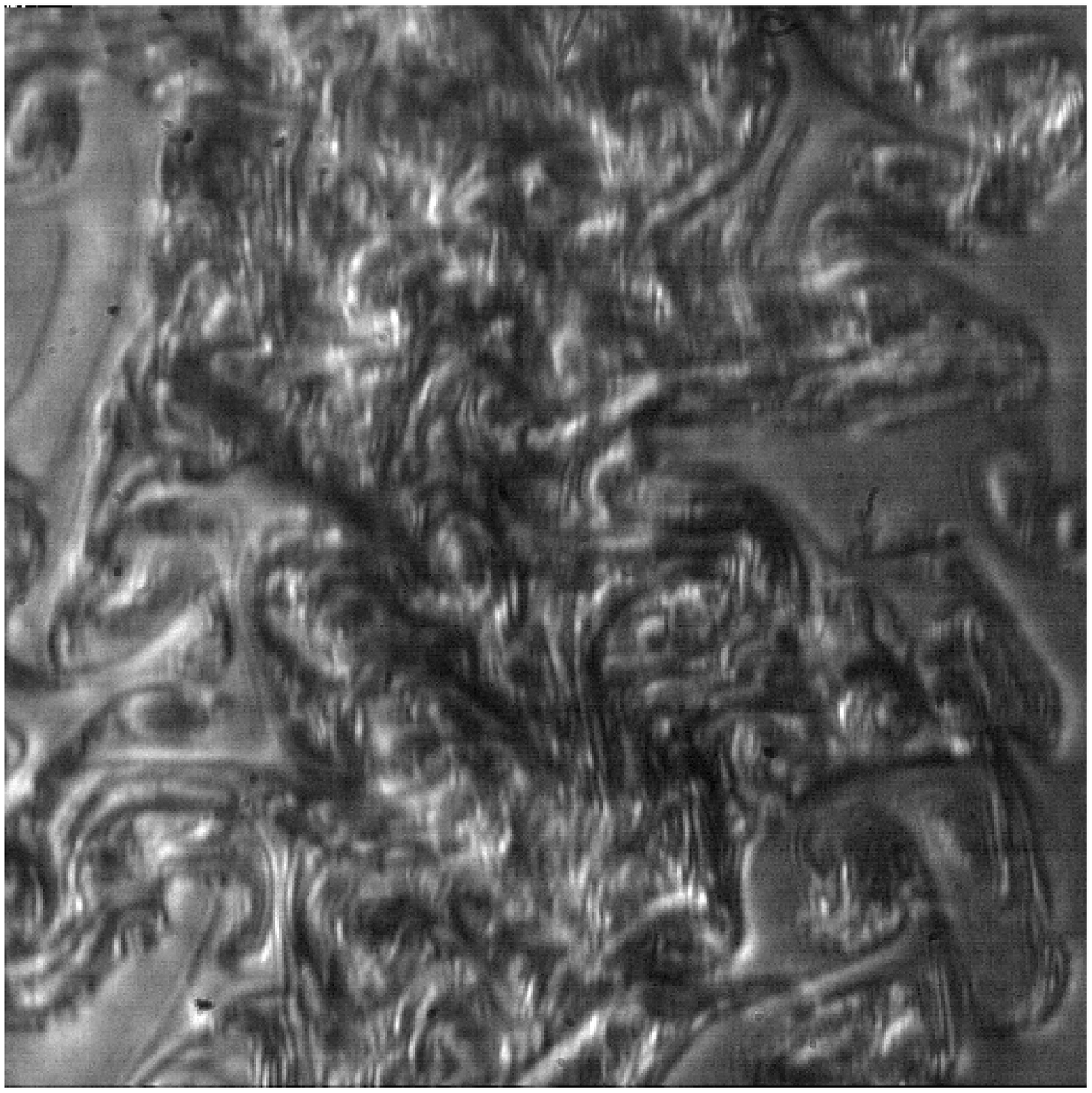}
\includegraphics[width=4.5cm]{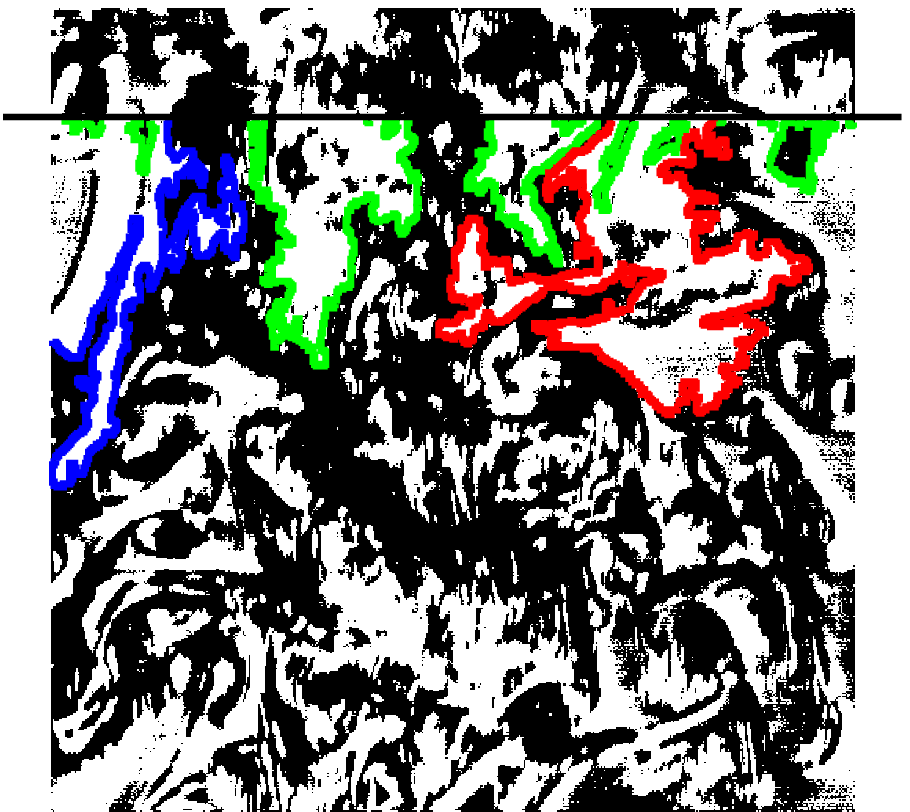}
\caption{(Color online) {\it Left:} Typical raw Schlieren image of the soap film for 
configuration A. {\it Right:} Extraction of paths with constant thickness. The shortest 
paths in light gray (green), whose lengths are below $l_\textrm{tr}$, are eliminated.
} \label{fig:image} \end{figure}

In this letter, we show evidence of conformal invariance in experiments of turbulence in 
soap films \cite{Couder84,Kellay98,Rivera98}. Such experiments approximate 2D turbulence, 
although deviations arise from variations in the soap film thickness, and also because 
of the friction in the presence of rough boundaries \cite{guttenberg}. Indirect vorticity 
measurements have been done \cite{Kellay98, Rivera98} and despite these effects, its 
dynamic is in some cases consistent with predictions for 2D turbulence. Also, to lowest 
order in the film thickness, thickness is advected as a passive scalar, and a strong 
correlation between thickness fluctuations and vorticity has been reported \cite{Rivera98}. 
However, both weak and strong intermittency effects have also been observed, in the direct 
and the inverse cascade, attributed to finite thickness effects \cite{Belmonte99}. The 
experiments we present here share these properties; a detailed description of PIV 
measurements is left for the future.

Given these similarities and differences, it is interesting to know whether conformal 
invariance can be identified in the system at hand. In the experiment, grid turbulence 
is generated using grids with different spacing of wires, while the fluid flows through 
a vertical channel as a result of gravity. To avoid indirect vorticity measurements and 
loss of resolution associated with PIV, we rely for the analysis on raw images of film 
thickness variation. The thickness variation in the film is visualized using a Schlieren 
technique \cite{Settles01}. When employing this technique, gradients of film thickness act 
on light rays as density gradients do in volumetric flows. Changes in the thickness act as 
changes in the refraction index, and the deviation of the light rays can be used to 
obtain an intensity image proportional to thickness variations. The use of this method 
allows us to have images with high contrast and resolution. Conformal invariance is then 
examined studying isolevels of constant film thickness. As an illustration, we present two 
experimental configurations: one with a grid size such that a wide inverse cascade of 
energy develops, and another with a grid size such that only a short range of inverse 
cascade is available. In the former case, evidence of conformal invariance is found, while 
in the latter conformal invariance is not observed.

\begin{figure}
\includegraphics[width=4.05cm]{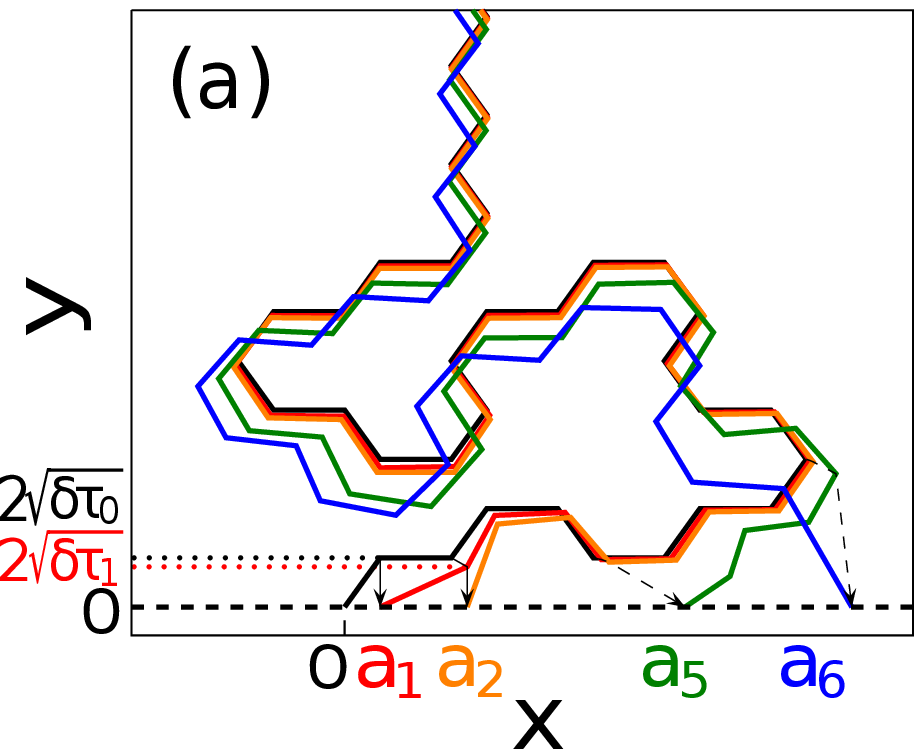}
\includegraphics[width=4.3cm]{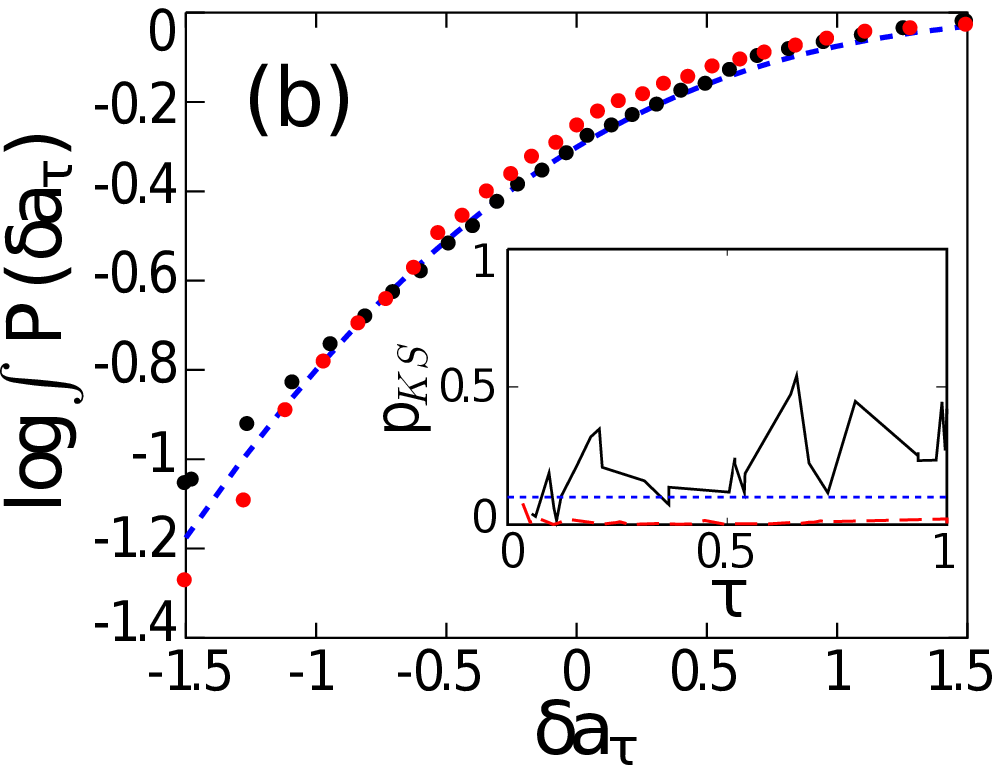}  \\
\vskip .2cm
\includegraphics[width=7.6cm]{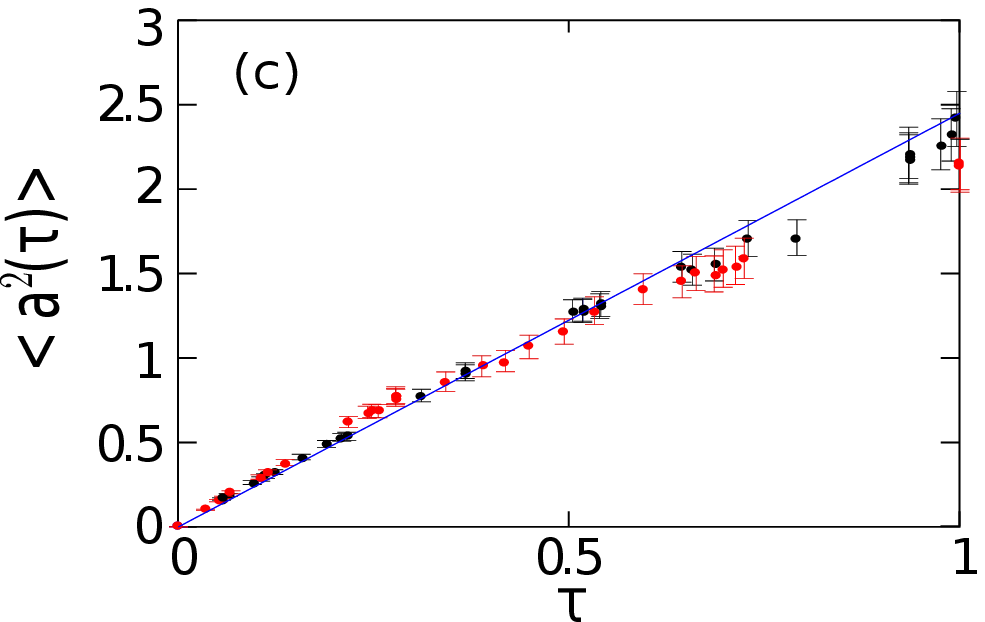}
\caption{(Color online) (a) The zipper algorithm: the original path (black) is gradually 
conformally squeezed down onto the $x$-axis (indicated by the gray paths and the dashed 
arrows). This defines a 1D path $a_n(\tau_n)$ (see \cite{cardy}). (b) Cumulative PDF of 
the increments $\delta a_\tau$, displayed here for $L = 800$ pixels for configurations A 
(black dots) and B (gray or red dots); the dashed (blue) line corresponds to a cumulative 
(normalized) Gaussian distribution. The inset shows the Kolmogorov-Smirnov probability 
$p_{KS}$, testing the null hypothesis that the increments fall into a Gaussian 
distribution. Black lines are for dataset A and long-dashed gray (red) lines for dataset 
B; the dashed (blue) line indicates a conventional threshold at $0.1$. (d) Scaling with 
error bars of the variance of $a$ with driving time $\tau$; the best fit for configuration 
A (black dots) yields $\kappa =2.5 \pm 0.3$ (solid blue line). Configuration B (gray or 
red dots) has a less clear scaling and does not pass the Gaussianity test.}
\label{fig:scaling1} 
\end{figure}

The experimental setup is as follows. We use a z-type Schlieren configuration with two 
mirrors of 50 cm in diameter and $3.0$ m focal length. Two razor blades perpendicularly 
disposed at the sagital and meridional focal planes enable us to obstruct the deviated 
light rays and to improve the contrast of the otherwise shadowgraph image that would form 
at the CCD of a monochromatic high-speed video camera (3000 fps with a resolution of 
$512\times 512$ pixels). Images are captured at a rate of 2500 fps, with one pixel 
corresponding to $0.161$ mm (corresponding to an area of $82.4 \times 82.4$ mm$^2$ 
centered in the middle of the channel width). The vertical soap film channel has a width of 
160 mm, and a length of $0.6$ m in the region of constant width (the expansion and the 
contraction of the tunnel have the same length). In the first configuration (configuration 
A), a grid with a separation between wires of $5.1$ mm and a width of $65$ mm is used to 
create turbulence downstream. The mean flow velocity is $1400$ mm/s, the mean film thickness 
is $5.6$ $\mu$m, and the Reynolds number (based on the injection scale) is $Re\approx 870$. 
When the third order structure function is computed from PIV data, a range with negative 
energy flux (corresponding to an inverse cascade) can be identified for scales between 
$\approx 30$ and 110 mm. In the second configuration (configuration B), a grid with a 
separation between wires of $0.3$ mm and a width of $45$ mm is used instead. The mean 
flow velocity is $1930$ mm/s, the channel width is $160$ mm, the mean film thicknes is 
$4.1$ $\mu$m, and the Reynolds number is $Re\approx 70$. Configuration B has an estimated 
inverse cascade range (from third order structure functions) between $\approx 25$ and 45 mm. 

Figure \ref{fig:image} shows a Schlieren image of the soap film. In each analysis, 2000 
snapshots of the film are used. An average of all snapshots is used as the background image, 
that we remove to undertake the analysis. The resulting image corresponds to thickness 
gradients, with zero value associated to constant thickness. For each image we then look at 
its four possible orientations, and for each one of those we set up an arbitrary $x$-axis 
(see the right panel in Fig~\ref{fig:image}). From each axis, we then explore all curves 
with no thickness variation until the curve goes out of the image or it returns to the 
$x$-axis. When the paths are not self-avoiding, we erase the loops similarly to how the 
loops are erased in loop erased random walks (LERWs) \cite{Lowner23}. Finally, only the 
paths whose length are above a given threshold $l_\textrm{tr}$ are kept. Here we consider 
$l_\text{tr} = 500$ pixels. We thus obtain a collection of paths which are correlated. In 
order to avoid redundancies (as the flow has a typical time-correlation given by its 
turnover time), we impose that the paths coming from two different snapshots have to be 
separated by a certain number of snapshots $n_\textrm{fil}$. The results given below are 
obtained with $n_\textrm{fil} = 8$, but we checked that $n_\textrm{fil} = 4$, 8, 12, 
16, and 20 give the same results. For $n_\textrm{fil} > 20$, similar results are obtained 
although with worse statistical accuracy as less curves are preserved 
($n_\textrm{fil} = 20$ roughly corresponds to the turnover time of the eddies at the 
injection scale in configuration A). With $n_\textrm{fil} = 8$, the number of paths above 
500 pixels is 1934 for configuration A, and 2216 for configuration B.

These curves are projected with a conformal map into 1D paths (the {\it driving functions}, 
which we will label $a(\tau)$), as done in \cite{Bernard06}. The zipper algorithm is used 
to obtain the 1D paths; see \cite{cardy} and Fig.~\ref{fig:scaling1}(a). The parametric 
variable $\tau$ (the ``time'') here only labels the successive points of the 1D paths, 
and should not be confused with the actual time corresponding to the different snapshots 
in the experiment. For increasing $\tau$, all curves in the 2D plane are generated in the 
direction that keeps positive vorticity to the right. Following Schramm's results 
\cite{Lowner23}, in order to check whether the measure of these curves is conformal 
invariant, we have to test whether these driving functions are likely to be seen as Brownian 
1D paths. In other words, the paths must be of the form $a(\tau) = \sqrt{\kappa} B(\tau)$, 
where $\kappa$ is a diffusion coefficient, and $B(\tau)$ is a standard Brownian motion. 
To test this, we first perform a Kolmogorov-Smirnov test, testing the null hypothesis that 
the increments of the driving functions fall into a Gaussian distribution (whose mean and 
variance are data driven). To compute the increments, a length $L$ is picked, and all 
driving functions with length larger than $L$ are selected. Then, the minimum driving time 
$\tau$ for which $L$ is reached in all these functions is found, and the increments are 
defined as $\delta a_\tau = a(\tau)$ with $\tau$ the minimum driving time for the given 
$L$. As an example, the cumulative probability density function (PDF) of the increments 
for datasets A and B and for $L=800$ pixels is shown in Fig.~\ref{fig:scaling1}(b). The 
$p_{KS}$ values of the Kolmogorov-Smirnov test for different minimum driving times $\tau$ 
(i.e., different lengths $L$) are shown in the inset of Fig.~\ref{fig:scaling1}(b). For 
configuration A the null hypothesis is rejected. Finally, we study the dependence of the 
variance of $a$ as a function of the driving time $\tau$ for datasets A and B to evaluate 
$\kappa$; see Fig.~\ref{fig:scaling1}(c). For configuration A we obtain 
$\kappa \approx 2.6$; configuration B presents slightly larger deviations from the linear 
behavior and does not pass the Gaussianity test.

\begin{figure}
\includegraphics[width=8.2cm]{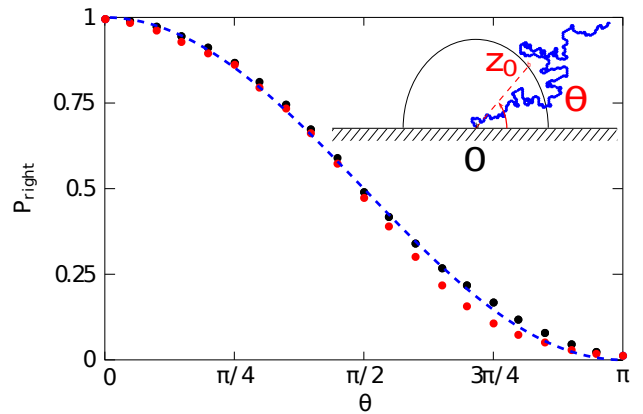}
\caption{(Color online) Probability that a path keeps a given point $z_0$ to its right. 
The dashed (blue) line indicates the SLE prediction given by Eq.~(\ref{proba}) with 
$\kappa = 8/3$. Black dots are for dataset A and gray (red) dots for dataset B. In the 
latter a larger departure is observed from the predicted behavior. The inset shows a 
sketch of a 2D path that keeps the given point $z_0$ to its left, and indicates how the 
angle $\theta=arg(z_0)$ is defined.}
\label{fig:pleft} 
\end{figure}

The procedure described above was also tested on synthetic 2D paths generated from LERWs, 
and from self avoiding walks (SAWs). The former are generated with a 2D Brownian motion 
in the upper plane whose loops are successively erased, while the latter are generated 
with a Markov chain Monte Carlo algorithm (the ``pivot algorithm'' \cite{Lal69}). In each 
case, 1000 paths are used. Both synthetic datasets pass the Gaussianity test and have 
variance $\kappa = 2.0 \pm 0.2$ (LERWs) and $2.9 \pm 0.3$ (SAWs).

The driving functions for configuration A show good $p_{KS}$ values as well as good 
scaling of the variance with $\tau$. The analysis thus indicates that the isolines used 
are good SLE candidates.  The value found is compatible with $\kappa = 8/3$, reported  
previously in numerical simulations \cite{Bernard06}, under the hypothesis that we are 
measuring here the properties of the envelope of the paths, thus accessing to the dual. 
This value corresponds to the continuum limit of SAWs. Configuration B, with a smaller 
Reynolds number and a narrower inverse energy cascade range, may not be sufficiently 
turbulent for conformal invariance to develop and be identified.

To confirm these findings, we now consider two properties that isocontours should exhibit 
if they are conformal invariant. First, we calculate the fractal dimension $d_f$ of each 
of the paths, and study the distribution of $\tilde{\kappa} = 1+d_f/8$. For the two 
synthetic datasets mentioned before, the distributions of $\tilde{\kappa}$ based on the 
fractality of the 2D paths have mean $1.9$ and variance $0.3$ in the LERWs, and 
$2.4 \pm 0.5$ in the SAWs. For dataset A, $\tilde{\kappa}=2.4 \pm 0.4$, while for dataset 
B, $\tilde{\kappa}=2.3\pm 0.6$. In other words, fractality (as given by a measure of 
$d_f$) obtains in both cases A and B, and we observe good agreement between $\kappa$ and 
the value derived from the distribution of the fractal dimensions of the 2D paths. 

Second, we verify a non-trivial result predicted by the SLE theory. If the curves are 
conformal invariant, then the probability that a path keeps a given point $z_0$ to its 
right depends only on $\kappa$ and $\theta = \arg(z_0)$, namely \cite{cardy}
\begin{equation}
p_\textrm{right} = \frac{1}{2} + \dfrac{\Gamma(\frac{4}{\kappa})}{\sqrt{\pi}\Gamma
    (\frac{8-\kappa}{2\kappa})} {{}_2 F_1}\left(\frac{1}{2},\frac{4}{\kappa},\frac{3}{2},
    -\cot^2 \theta\right) \cot \theta
\label{proba}  
\end{equation}
where $\Gamma$ is the Gamma function, and ${{}_2 F_1}$ is the Gauss hypergeometric 
function. This is a good example of non-trivial exact predictions that can be obtained 
if 2D or quasi-2D turbulence is conformal invariant (see \cite{Bernard06} for other 
examples). Results are shown in Fig.~\ref{fig:pleft}. Although both configurations 
show good agreement with the prediction, the agreement is better for dataset A.

The results indicate that the soap film flows studied here exhibit a behavior consistent 
with conformal invariance when an inverse energy cascade is sufficiently developed, at 
least to the point that SLE predictions are in good agreement with the experimental 
results. Note that an experimental result for SLE in a different context, that of the 
rugosity of tungsten oxide surfaces, has been reported recently \cite{saberi}, identified 
in their case with the Ising model ($\kappa=3$). Given the role played by 2D and quasi-2D 
flows as simple (although incomplete) models of atmospheric an oceanic turbulence, it is 
to be hoped that studies which bring together fluid dynamics and critical phenomena, 
together with the link to conformal field theory, will shed new light on the statistical 
properties of such flows.

{\it NCAR is sponsored by the National Science Foundation. PDM \& MIA acknowledge support 
from grants UBACYT 20020090200692 and PICT 2007-02211, GA \& MIA from grants PIP 3003, 
UBACYTI017, and PICT 12-9482, and AP and PDM from grant NSF-CMG 1025166DMS.}

\end{document}